\documentclass[a4paper, amsfonts, amssymb, amsmath, reprint, showkeys, nofootinbib, twoside, floatfix,onecolumn]{revtex4-1}
\usepackage[pdftex, pdftitle={Article}, pdfauthor={Author}]{hyperref} 
\usepackage{float}
\usepackage[english]{babel}
\usepackage[utf8]{inputenc}
\usepackage[colorinlistoftodos, color=green!40, prependcaption]{todonotes}
\usepackage{mathtools}
\usepackage{physics}
\usepackage{xcolor}
\usepackage{graphicx}
\usepackage[left=20mm, right=20mm, top=25mm, bottom =20mm, columnsep=15pt]{geometry} 
\usepackage{adjustbox}
\usepackage{placeins}
\usepackage[T1]{fontenc}
\graphicspath{ {./sections/} }

\begin{document}
\title{The reheating constraints to natural inflation in Horndeski gravity}
\author{Chen-Hsu Chien} \email[]{chenhsu0223@gmail.com}
\affiliation{Leung Center for Cosmology and Particle Astrophysics, National Taiwan University, Taipei 10617, Taiwan, ROC}
\author{Seoktae Koh} \email[]{kundol.koh@jejunu.ac.kr}
\author{Gansukh Tumurtushaa} \email[]{gansuh.mgl@gmail.com}
\affiliation{Department of Science Education, Jeju National University, Jeju, 63243, Korea}
\date{\today} 
\begin{abstract}
\label{abstract}
    For the subclass of Horndeski theory of gravity, we investigate the effects of reheating on the predictions of natural inflation. In the presence of derivative self-interaction of a scalar field and its kinetic coupling to the Einstein tensor, the gravitational friction to inflaton dynamics is enhanced. As a result, the tensor-to-scalar ratio $r$ is suppressed. We place the observational constraints on a natural inflation model and show that the model is now consistent with the observational data for some plausible range of the model parameter $\Delta$, mainly due to the suppressed tensor-to-scalar ratio. To be consistent with the data at the $1\sigma$ ($68\%$ confidence) level, a slightly longer  $N_k\gtrsim60$ duration of inflation than usually assumed is preferred. Since the duration of inflation, for any specific inflaton potential, is related to reheating parameters, including the duration $N_{re}$, temperature $T_{re}$, and equation-of-state $\omega_{re}$ parameter during reheating, we imposed the effects of reheating to the inflationary predictions to put further constraints. The results show that the duration of inflation $N_k$ is affected by considerations of reheating, mainly by the $\omega_{re}$ and $T_{re}$ parameters. If reheating occurs instantaneously for which $N_{re}=0$ and $\omega_{re}=1/3$, the duration of inflation is estimated to be $N_k\simeq57$, where the exact value is less sensitive to the model parameter $\Delta$ compatible with the CMB data. The duration of inflation is longer (or shorter) than $N_k\simeq57$ for the equation of state larger (or smaller) than 1/3 hence $N_{re}\neq0$. The maximum temperature at the end of reheating is $T_{re}^\text{max}\simeq3\times 10^{15}$ GeV, which corresponds to the instantaneous reheating. The low reheating temperature, as low as a few MeV, is also possible when $\omega_{re}$ is closer to $1/3$.     
\end{abstract}
\maketitle
\section{Introduction} \label{sec:intro}
   Cosmic inflation is a period of accelerated expansion of the early universe and an attractive mechanism for explaining the observed structures in the universe. It provides a solution to several puzzles, so-called cosmological problems, including the horizon, flatness, and monopole problems~\cite{Guth:1980zm, Linde:1981mu, Linde:1982zj, Albrecht:1982wi, Linde:1983gd}. Current observations favor the inflationary paradigm for its prediction of an almost scale-invariant spectrum of primordial curvature perturbations imprinted in the Cosmic Microwave Background (CMB) radiation and the large scale structure formations data~\cite{Boomerang:2000efg, SDSS:2003eyi,SDSS:2004kqt,Blake:2011en}. The simplest model of inflation describes the period of exponential expansion of the early universe being driven by the slow-roll of a single scalar field known as the \emph{inflaton}. To match the CMB temperature anisotropy measurements and give rise to a sufficient amount of inflation, the models of inflation share the flatness of the inflaton potential as a common feature among them~\cite{Planck:2018jri}. This nontrivial requirement of flatness of inflaton potential creates fine-tuning problems, in particular, quantum corrections in the absence of a symmetry generically spoil the flatness of the potential, which is known as the $\eta$-problem.
   
   In particle physics model of inflation, the flatness of the potential should be protected against radiative corrections that can arise from the inflaton self-interactions or from its coupling to matter fields, responsible for reheating the universe after inflation. To protect the flatness of the inflaton potential against the radiative corrections, the symmetries of a system play an important role$-$\emph{e.g.,} as for the standard model Higgs field, supersymmetry can provide a natural protection against the radiative corrections. If the inflaton is a pseudo-scalar axion, the axionic (shift) symmetry, $\phi\to\phi+\text{constant}$, is another possible symmetry that protects the flatness of the inflaton potential. In this case, the inflaton potential arises from the breaking of a (global) shift symmetry, and the couplings of inflaton to matter fields do not affect the inflaton potential as long as the shift symmetry is respected. This mechanism was originally proposed in Ref.~\cite{Freese:1990rb} as \emph{natural inflation}, and its implications in light of the observational data, as well as implementations in the other theories, have been discussed in Ref.~\cite{ Adams:1992bn, Freese:2004un, Arkani-Hamed:2003xts, Kawasaki:2000yn, Kaplan:2003aj, Savage:2006tr, Freese:2014nla, Firouzjahi:2003zy, Hsu:2004hi, Freese:1994fp, Kim:2004rp}. In the natural inflation model, a rolling inflaton has a flat potential due to shift symmetries. Whenever a global symmetry is spontaneously broken, Nambu Goldstone bosons arise, with a potential that is exactly flat. However, the inflaton cannot roll and drive inflation if the shift symmetry is exact. Thus, to drive inflation, there must be additional explicit symmetry breaking hence these particles become pseudo-Nambu Goldstone bosons, with “nearly” flat potentials exactly as required by inflation~\cite{Freese:2004un}. The resulting inflaton potential is generally of the form $V(\phi)=\Lambda^4[1+\cos(\phi/f)]$, where $f$ is called the decay constant and is constrained by the observational data~\cite{Freese:1990rb,Adams:1992bn,Freese:2004un}. Despite its well established theoretical motivation and a simple form of the potential, the natural inflation model is disfavored at greater than 95\% confidence by the observational data~\cite{Planck:2018jri, Stein:2021uge, Huang:2015cke}, especially after the recent announcement of the improved Planck+BICEP/Keck 2018 data~\cite{BICEP:2021xfz}. Although the original model, in which the scalar field is minimally coupled to gravity, is now disfavored by current observational constraints, some variants of natural inflation, including the generalized and multi-field versions, were proposed and discussed in Refs.~\cite{Munoz:2014eqa, Zhang:2018wbn, Kim:2004rp, Civiletti:2020fkm, Forconi:2021que}, and references therein. The inflationary predictions in those extended models can be modified and, in general, be consistent with observations. 

   Adopting the original natural inflation potential without any modification$-$\emph{e.g.}, $V(\phi)=\Lambda^4[1+\cos(\phi/f)]$, we investigate natural inflation for the cosmological models with the derivative self-interaction of the inflaton field and inflaton's kinetic coupling to gravity via the Einstein tensor. 
   The cosmological model of our interest was proposed in Ref.~\cite{Tumurtushaa:2019bmc} (see Refs.~\cite{Bayarsaikhan:2020jww, Chen:2021nio} for its applications) as a subclass of $G-$inflation framework~\cite{Kobayashi:2011nu}, which is based on the Horndeski (equivalently,  generalized Galileon) theory~\cite{Horndeski:1974wa, Deffayet:2011gz}, \emph{i.e.}, the most general scalar-tensor theory with second-order field equations avoiding the Ostrogradski instability~\cite{Ostrogradski}, see Ref.~\cite{Kobayashi:2019hrl} for a review. A unique feature of considering inflaton's derivative self-interaction and its kinetic coupling with gravity is that the gravitationally enhanced friction mechanism works for steep potentials~\cite{Germani:2010gm, Tsujikawa:2012mk, Tsujikawa:2013ila, Germani:2011ua, Kamada:2010qe}.
   Thus, it motivates us to investigate whether it is possible to reconcile the predictions of natural inflation with the CMB observations in the setting of inflaton's derivative self-interaction and its kinetic coupling with gravity scenario. 
   
   After inflation, it is usually assumed that the inflaton coherently oscillates at the minimum of its potential, decaying and transferring its energy to a relativistic plasma of the standard model particles~\cite{Abbott:1982hn,Dolgov:1982th,Albrecht:1982mp}. This post-inflationary process that populates our universe with ordinary matter is known as \emph{reheating}, see Ref.~\cite{Allahverdi:2010xz} for a review. Because there are no direct cosmological observations are traceable from reheating, the physics of reheating is highly uncertain and unconstrained. Thus, this post-inflationary era depends heavily on models of inflation. However, it was pointed out in the literature that consideration of reheating may provide additional constraints to inflationary perdictions~\cite{Martin:2010kz, Adshead:2010mc, Mielczarek:2010ag, Dodelson:2003vq, Easther:2011yq}. In this work, we follow the approaches proposed in Ref.~\cite{Martin:2010kz, Adshead:2010mc, Mielczarek:2010ag, Dodelson:2003vq, Easther:2011yq, Dai:2014jja,Munoz:2014eqa,Creminelli:2014fca, Cai:2015soa} to perform the analyses on the reheating parameters, including the equation of state, the duration, and the temperature of reheating.  Then, using the link between these parameters of reheating and the observable quantities of inflation, we provide constraints on the inflationary predictions of natural inflation in light of current observational data~\cite{Planck:2018jri, BICEP:2021xfz}.
   
   This paper is organized as follows. We start Sec.~\ref{sec:bg-pert} by setting up our model with the derivative self-interaction of the scalar field and the kinetic coupling between the scalar field and gravity, which belongs to the subclass of Horndeski's theory of gravity. Then, in the same section, we derive the background equations of motion, as well as the observable quantities; including power spectra of tensor and scalar modes, $\mathcal{P}_S$ and $\mathcal{P}_T$, respectively, their spectral tilts $n_S$ and $n_T$, and the tensor-to-scalar ratio $r$ in the slow-roll scenario of inflation. Based on our analytic results obtained in Sec.~\ref{sec:bg-pert}, we place the observational constraints on $n_S$ and $r$ predictions of natural inflation in Sec.~\ref{sec:obs-con} and show that the natural inflation model is now consistent with the latest observational data, mainly due to the suppressed tensor-to-scalar ratio. In Sec.~\ref{sec:reheating}, we impose the effects of reheating to the predictions of natural inflation. The results of the section implies that the reheating parameters, the equation of state $\omega_{re}$ and the temperature $T_{re}$ at the end of reheating, significantly affect the duration of inflation $N_k$ hence the $n_S$ and $r$. We conclude our work and provide some implications of our findings in Sec.\ref{sec:conclusions}.
      
\section{Background and Perturbation dynamics} \label{sec:bg-pert}
The action for the cosmological model that we investigate in this work is give as~\cite{Tumurtushaa:2019bmc, Bayarsaikhan:2020jww, Chen:2021nio}
\begin{align}\label{eq:action}
    S =& \int d^{4}x\sqrt{-g} \left[ \frac{M_p^2}{2} R - \frac{1}{2}
    \left(g^{\mu\nu}-\frac{\alpha}{M^3}g^{\mu\nu}\partial_\rho\partial^\rho\phi + \frac{\beta}{M^2} G^{\mu\nu}\right)\partial_\mu\phi\partial_\nu \phi - V(\phi)  \right]\,,
\end{align}   
where $M_p=2.44\times 10^{18}$GeV is the reduced Planck mass and $V(\phi)$ is the inflaton potential, and $M$ is the mass scale making $\alpha$ and $\beta$ dimensionless constants. In the limit $(\alpha, \beta)\rightarrow 0$, the Einstein gravity with a minimal coupling to the scalar field is recovered. Thus, the case with $(\alpha, \beta)\neq0$ reflects the deviation from the general relativity. While Refs.~\cite{Tumurtushaa:2019bmc, Bayarsaikhan:2020jww, Chen:2021nio} discuss the scalar field dependent coupling function $\xi(\phi)$ for the $\alpha-$term in Eq.~(\ref{eq:action}), we consider such the coupling to be a constant; namely, $\xi(\phi)=1$, in the present study for simplicity. 

Varying this action with respect to metric $g_{\mu\nu}$, one obtains the Einstein equation 
\begin{eqnarray}\label{eq:Einsteineq}
    G_{\mu\nu}= \frac{1}{M_p^2} T_{\mu\nu}\,,
\end{eqnarray}
where the $G_{\mu\nu}=R_{\mu\nu}-g_{\mu\nu}R/2$ is the Einstein tensor and the energy momentum tensor is 
\begin{align}\label{emt}
    T_{\mu\nu} =&\partial_\mu\phi \partial_\nu\phi-\frac{1}{2}g_{\mu\nu}
    \left(\partial_\alpha\phi\partial^\alpha\phi+2V\right)+\frac{\alpha}{M^3}\left[(\nabla_\alpha\phi\nabla^\alpha\phi)_{(\mu}\nabla_{\nu)}\phi-\square\phi \nabla_\mu\phi\nabla_\nu\phi
    -\frac{1}{2}g_{\mu\nu}(\nabla_\alpha\phi\nabla^\alpha\phi)_{\beta}\nabla^\beta\phi\right]\nonumber\\
    &+\frac{\beta}{M^2} \left[-\frac{1}{2}\nabla_\mu\phi \nabla_\nu\phi R + 2\nabla_\alpha\phi \nabla_{(\mu}\phi R^{\alpha}\,_{\nu)}+\nabla^\alpha\phi\nabla^\beta\phi R_{\mu\alpha\nu\beta}+\nabla_\mu\nabla^\alpha\phi\nabla_\nu\nabla_\alpha\phi-\nabla_\mu\nabla_\nu\phi\square\phi\right.\\
    &-\frac12 G_{\mu\nu} \nabla_\alpha\phi \nabla^\alpha\phi+\frac{1}{2}g_{\mu\nu}
    \left.\left(\left(\square\phi\right)^2-\nabla^\alpha\nabla^\beta\phi\nabla_\alpha\nabla_\beta\phi-2\nabla_\alpha\phi\nabla_\beta\phi R^{\alpha\beta}\right)\right]\,. \nonumber
\end{align}
In addition, from Eq.~(\ref{eq:Einsteineq}) using the Bianchi identity $\nabla^\mu G_{\mu\nu}=0$, we obtain the evolution equation for the scalar field as
\begin{align}
    \nabla^{\mu} T_{\mu\nu} = 0\,.
\end{align}

In a spatially flat Friedman-Robertson-Walker universe with metric
\begin{eqnarray}\label{eq:flatMetric}
ds^2=-dt^2+a(t)^2\delta_{ij}dx^i dx^j\,,
\end{eqnarray} 
where $a(t)$ is a scale factor, the background and field equations are obtained as
\begin{eqnarray}
&& 3M_p^2H^2 = \frac{1}{2} \dot{\phi}^2 +V(\phi)  +\frac{3\alpha}{M^3}H\dot{\phi}^3 -\frac{9\beta}{2M^2}\dot{\phi}^2 H^2 
\,,\label{eq:EE00}\\
&& M_p^2\left( 2\dot{H}+3H^2 \right) = -\frac{1}{2}\dot{\phi}^2 +V  +\frac{\alpha}{M^3}\dot{\phi}^2\ddot{\phi} -\frac{\beta}{2M^2}\dot{\phi}^2\left(2\dot{H} +3H^2 +4H\frac{\ddot{\phi}}{\dot{\phi}} \right)
\,,\\
&& \ddot{\phi} +3H\dot{\phi} + V_{,\phi} +\frac{3\alpha}{M^3}\dot{\phi}^2 \left( \dot{H} + 3H^2 + 2 H \frac{\ddot{\phi}}{\dot{\phi}} \right)-\frac{3\beta}{M^2} H \dot{\phi}\left( 2\dot{H} +3H^2 +H\frac{\ddot{\phi}}{\dot{\phi}}\right) =0\,,\label{eq:fieldEq}
\end{eqnarray}
where the overdot denotes the derivative with respect to time $t$ and $V_{,\phi}\equiv dV(\phi)/d\phi$. To investigate the slow-roll inflation, we take so called the slow-roll conditions that read $V(\phi)\gg \dot{\phi}^2$ and $\ddot{\phi}\ll 3H\dot{\phi}$ into account and introduce the following slow-roll parameters
\begin{eqnarray}\label{eq:srparams}
\epsilon_1 \equiv -\frac{\dot{H}}{H^2}\,,\quad 
\epsilon_2 \equiv -\frac{\ddot{\phi}}{H\dot{\phi}}\,,\quad 
\epsilon_3 \equiv \frac{\dot{\phi}^2}{2M_p^2H^2}\,, 
\end{eqnarray}
which assumed to be small, $\epsilon_{i}\ll1$ where $i=1,2,3$, during inflation. Eq.~(\ref{eq:fieldEq}) can, therefore, be rewritten in terms of these parameters as
\begin{eqnarray}\label{eq:fieldEqSR}
3H\dot{\phi}\left[1-\frac{1}{3}\epsilon_2 +\frac{3\alpha}{M^3} H\dot{\phi}\left( 1-\frac{\epsilon_1}{3}-\frac{2\epsilon_2}{3}\right)-\frac{3\beta}{M^2} H^2\left(1-\frac{2\epsilon_1}{3}-\frac{\epsilon_2}{3}\right)  \right] = -V_{,\phi}\,.
\end{eqnarray}
Thus, under the slow-roll conditions, Eqs.~(\ref{eq:EE00}) and (\ref{eq:fieldEq}) can be approximated as 
\begin{eqnarray}\label{eq:MFEq}
    3M_p^2H^2\simeq  V\,, \quad 3H\dot{\phi} \left( 1+\mathcal{A}\right) \simeq -V_{,\phi} \,,
\end{eqnarray} 
where 
\begin{align}\label{eq:curA}
    \mathcal{A}=\frac{3\alpha}{M^3} H\dot{\phi}-\frac{3\beta}{M^2} H^2\,.
\end{align}
When $\mathcal{A}\ll1$, Eq.~(\ref{eq:MFEq}) describes the standard slow-roll inflation in the Einstein gravity. Thus, the deviation from GR implies $\mathcal{A}\gtrsim 1$. In addition to Eq.~(\ref{eq:srparams}), we introduce the following parameter
\begin{align}\label{eq:gamma}
    \gamma\equiv \frac{\alpha}{\beta M} \frac{ \dot{\phi}}{H}\,,
\end{align}
to weigh the contributions of each term in Eq~(\ref{eq:curA}). The implication of introducing $\gamma$ is as follows: if the kinetic coupling between the scalar field and gravity is much stronger (or weaker) than the derivative self-interaction of the scalar field, then we get $|\gamma|\ll1$ (or $|\gamma|\gg1$). The both terms contribute equally during inflation for $|\gamma|\sim\mathcal{O}(1)$. 
Consequently, from Eq.~(\ref{eq:curA}), $\mathcal{A}\gtrsim1$ implies
\begin{align}
    \frac{H^2}{M^2}\gtrsim \frac{1}{3\beta(\gamma-1)}\,, 
\end{align}
where $\beta(\gamma-1)\neq0$ is assumed. If $\gamma=1$ when it is evaluated some time during or at the end of inflation, which indicates that the additional $\alpha$ and $\beta$ terms in Eq.~(\ref{eq:curA}) cancel each other at that time. Thus, without loss of generality, we consider $\gamma\neq1$ to have nozero effects of these additional contributions. 
The amount of inflation is quantified by the number $N_k$ of $e$-folds, which reads
\begin{align}\label{eq:efold}
    N_k = \int_\phi^{\phi_e} \frac{H}{\dot{\tilde{\phi}}} d\tilde{\phi} \simeq \frac{1}{M_p^2}\int_{\phi_e}^\phi \frac{V}{V_{,\tilde{\phi}}}\left( 1+\mathcal{A}\right)d\tilde{\phi}\,,
\end{align}
where $\phi_e$ is the scalar-field value at the end of inflation and is to be estimated by solving $\epsilon_1(\phi_e)\equiv1$.

The linear perturbation analysis for our model Eq.~(\ref{eq:action}) is carried out in Ref.~\cite{Tumurtushaa:2019bmc, Chen:2021nio}, see Ref.~\cite{Kobayashi:2011nu} for more general cases,  and the observable quantities, including power spectra of the scalar ($\mathcal{P}_S$) and tensor ($\mathcal{P}_T$) modes, their spectral tilts ($n_S$ and $n_T$, respectively), and the tensor-to-scalar ratio $r$, are computed on the large scale $c_S k |\tau|\ll 1$ as
\begin{align}\label{eq:powerS}
    \mathcal{P}_S &= \frac{k^3}{2\pi^2}\left| \frac{\nu_k}{z_S}\right|^2 \simeq \frac{H^2}{8\pi^2 M_p^2 c_S^3 \epsilon_V}\left(1+\mathcal{A}\right)\,,
    \\
    \mathcal{P}_T &= \frac{k^3}{\pi^2}\Sigma_{\lambda=x, +}\left| \frac{\mu_{\lambda,k}}{z_T}\right|^2 \simeq \frac{H^2}{2\pi^2 M_p^2 c_T^3}\,, \label{eq:powerT}\\
    n_S-1 &= \frac{d\ln \mathcal{P}_S}{d\ln k} = 3-2\mu_S \simeq \frac{1}{1+\mathcal{A}} \left[2\eta_V -2\epsilon_V\left(4- \frac{1}{1+\mathcal{A}}\right) \right]\,,\label{eq:nsofphi}\\
    n_T &= \frac{d\ln \mathcal{P}_T}{d\ln k} = 3-2\mu_T 
    \simeq -\frac{2\epsilon_V}{1+\mathcal{A}}\,,\\
    r&= \frac{\mathcal{P}_T}{\mathcal{P}_S} \simeq \frac{16\epsilon_V}{1+\mathcal{A}}\,,\label{eq:rofphi}
\end{align}
where $\epsilon_V \equiv M_p^2\left(V_{,\phi}/V\right)^2/2$ and $\eta_V\equiv M_p (V_{,\phi\phi}/V)$, and $c_{S, T}^2\sim 1+\mathcal{O}(\epsilon_1)\simeq 1$, see Ref.~\cite{Tumurtushaa:2019bmc, Chen:2021nio} for further details. Based on these findings, we place observational constraints on the natural inflation model in the following sections. In the standard single field inflation scenario in the general relativity, which is $\mathcal{A}\ll1$ limit of our model, the theoretical predictions of natural inflation on $n_S$ and $r$ has been disfavored by observational data~\cite{BICEP:2021xfz, Planck:2018jri}. Thus, in the following section, we place observational constraints on natural inflation in the $\mathcal{A}\gg 1$ limit and show that the model is now consistent with the current observations even at $1\sigma$ ($68\%$ confidence) level. 

\section{Observational constraints}\label{sec:obsNI}
\label{sec:obs-con}
\begin{figure}[t]
    \centering
    \includegraphics[width=0.45\textwidth]{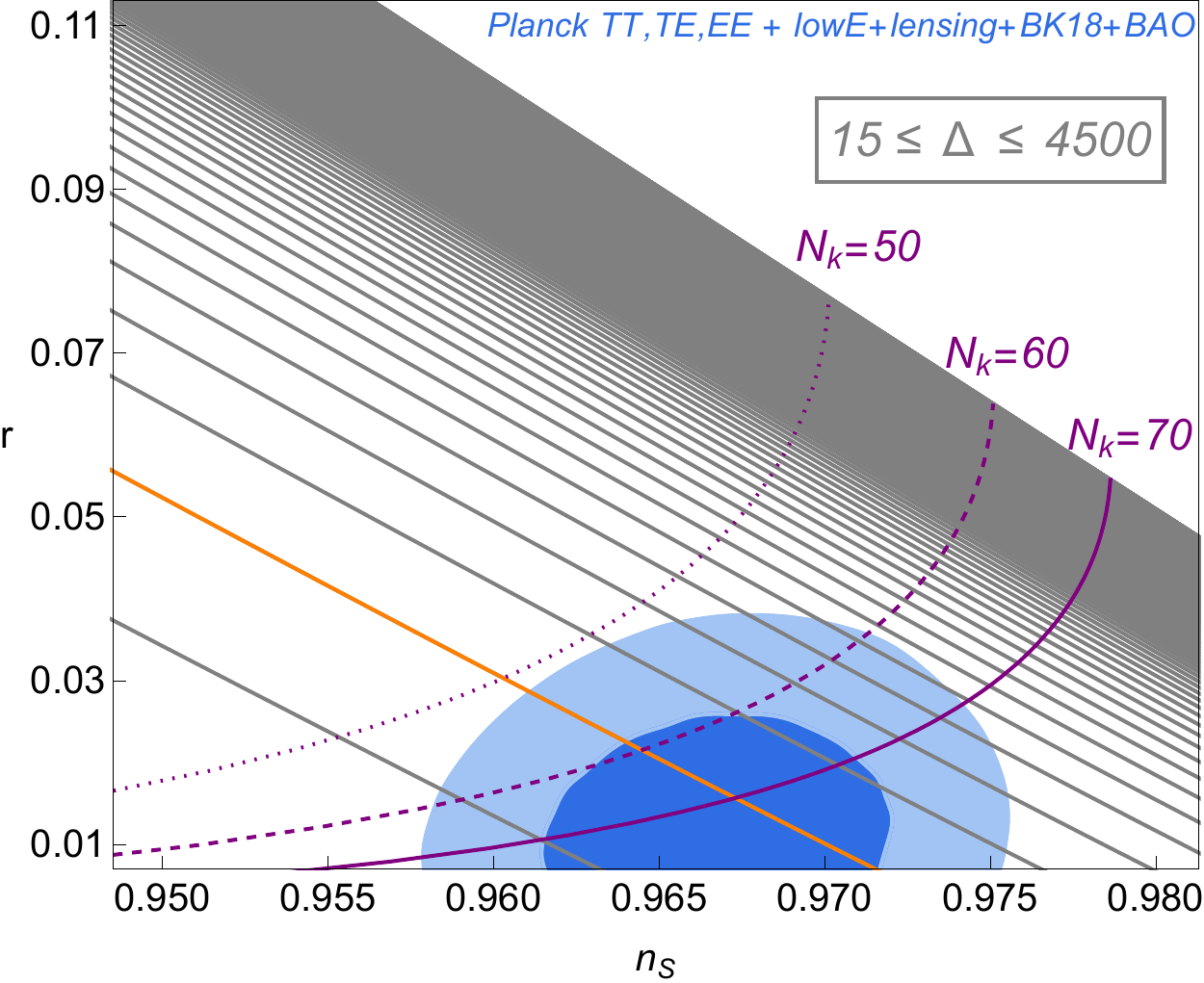}\hspace{0.5 cm}
    \includegraphics[width=0.45\textwidth]{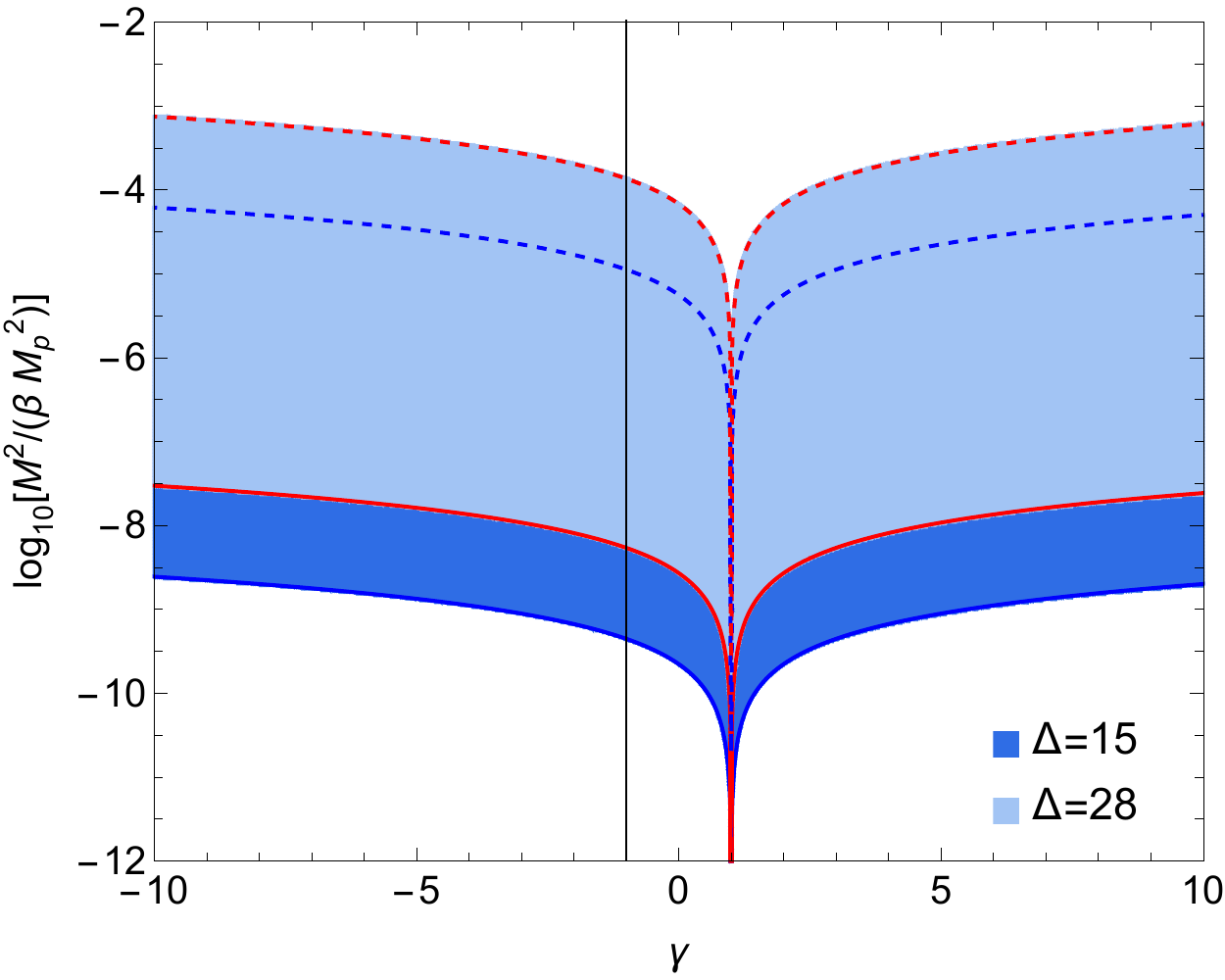}
    \caption{Left: The $n_S-r$ plot for natural inflation with potential given in Eq.~(\ref{eq:potential}). The shaded regions present the $1\sigma$ (dark blue) and $2\sigma$ (light blue) contours of the latest Planck+BICEP/Keck 2018 data~\cite{BICEP:2021xfz}. The diagonal gray lines indicate the different values of $\Delta$ within the interval $15\leq \Delta\leq 4500$, increasing from bottom left to top right with a step of $10$. For the orange line, $\Delta=20$. The purple lines represent the number $N_k$ of $e$-folds during inflation; $N_k=50$(dotted), $60$(dashed), and $70$(solid). Right: The parameter space of $\gamma-M$ from Eq.~(\ref{eq:Delta}), where $0.3<\log_{10}(f/M_p) < 2.5$ from Planck data~\cite{Planck:2018jri} is used. The shaded blue regions correspond to different values of $\Delta$, \emph{i.e.}, $\Delta=15$ (darker) and $\Delta=28$ (lighter). The boundaries of these shaded regions indicate the upper and lower limits of $0.3\leq \log_{10}(f/M_p)\leq 2.5$; namely, $\log_{10}(f/M_p)=0.3$ for the red- and blue-solid lines while $\log_{10}(f/M_p)=2.5$ for the red- and blue-dashed lines, respectively.} 
    \label{fig:nsvsr}
\end{figure}

In the presence of the derivative self-interaction of the scalar field and its kinetic coupling with the Einstein tensor, we place observational constraints on natural inflation. From now on, let us consider the case of $\mathcal{A}\gg1$.
The potential for natural inflation reads 
\begin{align}\label{eq:potential}
    V(\phi) = \Lambda^4 \left[ 1+ \cos\left(\frac{\phi}{f} \right)\right]\,,
\end{align}
where the energy density $\Lambda^4$ and the decay constant $f$ are the parameters of the model with dimensions of mass. In the limit $f\rightarrow\infty$, the potential behaves like pure power laws, \emph{e.g.}, $V(\phi)\sim m^2\phi^2$ where $m$ is an energy scale that plays the role of $\Lambda$~\cite{Munoz:2014eqa}.  For natural inflation, the number $N_k$ of $e$-fold is obtained from Eq.~(\ref{eq:efold}) as
\begin{align}\label{eq:efold2}
    N_k = \Delta \left[ \mathcal{F}\left(\phi_e\right) - \mathcal{F}\left(\phi\right) \right]\,,
\end{align}
where 
\begin{align}\label{eq:Delta}
    \Delta = \beta(\gamma-1) \frac{f^2 \Lambda^4}{M^2 M_p^4}\,,
\end{align}
and 
\begin{align}\label{eq:fend}
    \mathcal{F}\left(\phi\right) = \cos{\left(\frac{\phi}{f}\right)} + 4\ln{\sqrt{\frac{1}{2} \left[1-\cos{\left(\frac{\phi}{f}\right)}\right]}}\,.
\end{align}
In Eq.~(\ref{eq:Delta}), $\Delta$ becomes zero if $\gamma=1$, which indicates the complete absence of $\mathcal{A}$ in our model. Thus, we consider $\gamma\neq1$ case in this study. 
By using the condition that $\epsilon_1(\phi_e)=1$, we obtain the scalar-field value $\phi_e$ at the end of inflation as
\begin{align}\label{eq:cosphiend}
    \cos{\left(\frac{\phi_e}{f}\right)} = \frac{4}{1+\sqrt{16\Delta+1}}-1 \,.
\end{align}
Substituting Eq.~(\ref{eq:cosphiend}) into Eq.~(\ref{eq:fend}), we obtain
\begin{align}
    \mathcal{F}\left(\phi_e\right) = 2\ln \left[\frac{1+8\Delta-\sqrt{16\Delta+1}}{8\Delta}\right]-\frac{1+4\Delta-\sqrt{16\Delta+1}}{4\Delta}\,.
\end{align}
Consequently, the potential value at the end of inflation gets
\begin{align}\label{eq:potphie}
    V(\phi_{e}) = \left(\frac{\sqrt{16\Delta+1}-1}{4\Delta}\right)\Lambda^4\,.
\end{align}
For the potential in Eq.~(\ref{eq:potential}), the spectral index for scalar modes and the tensor-to-scalar ratio are obtained from Eqs.~(\ref{eq:nsofphi}) and (\ref{eq:rofphi}) as
\begin{align}\label{eq:nspred}
    n_S&=1 - \frac{2\left[2-\cos{\left(\frac{\phi}{f}\right)} \right]}{\Delta\left[ 1+\cos{\left(\frac{\phi}{f}\right)}\right]^2} \,,\\
    r&=\frac{8\left[1-\cos{\left(\frac{\phi}{f}\right)} \right]}{\Delta\left[ 1+\cos{\left(\frac{\phi}{f}\right)}\right]^2}\,.\label{eq:rpred}
\end{align}
Using Eq.~(\ref{eq:efold2}), we obtain
\begin{align}\label{eq:cosphi}
    \cos{\left(\frac{\phi}{f}\right)} = 1+ 2\mathcal{W}\left(\pm e^{\frac{1}{2}\left(\mathcal{F}\left(\phi_e\right)-\frac{N_k+\Delta}{\Delta}\right)} \right)\,,
\end{align}
where $\mathcal{W}(x)$ is the product logarithm function, also called the Lambert $\mathcal{W}$ function. 
Although we have several free parameters; including $\alpha$, $\beta$, and $M$, coming from our model, and $\Lambda$ and $f$, from the choice of inflaton potential, our result shows that the theoretical predictions of $n_S$ and $r$ depend only on a single parameter $\Delta$, which is given in Eq.~(\ref{eq:Delta}). Thus, it is effectively one parameter model, and we plot in the left panel of Fig.~\ref{fig:nsvsr} the theoretical predictions of our model based on Eqs.~(\ref{eq:nspred}) and (\ref{eq:rpred}) for natural inflation. 
The shaded dark- and light-blue regions represent the $1\sigma$ and $2\sigma$ confidence contours of the Planck+BICEP/Keck 2018 data~\cite{BICEP:2021xfz}, respectively. The diagonal gray lines indicate the different values of $\Delta$; for example, $\Delta=20$ for the orange line. The purple lines represent the number $N_k$ of $e$-folds during inflation; $N_k=50$(dotted), $60$(dashed), and $70$(solid). The values of $r$ and $n_S$ decrease/increase along the purple lines as the $\Delta$ value decreases/increases. The figure shows that natural inflation is now compatible with the observational data if the kinetic coupling between gravity and the scalar field and the derivative self-interaction of the scalar field are taken into account during inflation. For our model to be consistent with the observational data at $1\sigma$/$2\sigma$ level, the duration of inflation is preferred to be $N_k\gtrsim60$/$N
_k\gtrsim50$. The ranges of $\Delta$ compatible with the data are $18\lesssim\Delta\lesssim25$ ($1\sigma$ confidence) and $25\lesssim \Delta \lesssim 45$ ($2\sigma$ confidence) for $N_k=60$(dashed line), the most conservative value for duration of inflation. Having obtained the range of $\Delta$ values in agreement with the observational data, we can search for the parameter space of other parameters using Eq.~(\ref{eq:Delta}). 

The amplitude of the scalar power spectrum is well determined to be $\mathcal{P}_S(k_\ast)=2.0989\times10^{-9}$(TT, TE, EE+lowE+lensing) at the pivot scale $k_\ast=0.05\text{Mpc}^{-1}$~\cite{Planck:2018jri}. Thus, we can use Eq.~(\ref{eq:powerS}) to obtain the $\Lambda$ parameter. First, we write Eq.~(\ref{eq:powerS}) in the following form 
\begin{align}\label{eq:powerS01}
    \mathcal{P}_S(k_\ast) = \frac{\Delta}{12 \pi^2}\frac{\Lambda^4}{M_p^4} \left[1+\cos\left(\frac{\phi}{f}\right) \right]^3\left[1-\cos\left(\frac{\phi}{f}\right)\right]^{-1}=2.0989\times10^{-9}\,,
\end{align}
where $\cos(\phi/f)$ is given in Eq.~(\ref{eq:cosphi}) in terms of $N_k$ and  $\Delta$. The $N_k$ can be expressed in terms of $n_S$ and $\Delta$ after substituting Eq.~(\ref{eq:cosphi}) into Eq.~(\ref{eq:nspred}), which then reads
{\small \begin{align}\label{eq:Nkns}
    N_k = & \Delta \left\{\mathcal{F}(\phi_e) - \ln \left[ \exp\left(\frac{1-(n_S-1)\Delta+\sqrt{1-6(n_S-1)\Delta}}{(n_S-1)\Delta}\right)
    \left(\frac{1-2(n_S-1)\Delta+\sqrt{1-6(n_S-1)\Delta}}{2(n_S-1)\Delta}\right)^2 \right] \right\}\,.
\end{align}}
Thus, we obtain $\Lambda(n_S, \Delta)$ from Eq.~(\ref{eq:powerS01}) as
\begin{align}\label{eq:Lambda}
    \frac{\Lambda^4}{M_p^4} = -\frac{3\pi^2}{\Delta}\mathcal{P}_S(k_\ast) \left(1+ \mathcal{W}\left[-\sqrt{\exp\left(\mathcal{F}\left(\phi_e\right)-\frac{N_k+\Delta}{\Delta}\right)}\right] \right)^{-3} \mathcal{W}\left[-\sqrt{\exp\left(\mathcal{F}\left(\phi_e\right)-\frac{N_k+\Delta}{\Delta}\right)}\right]\,.
\end{align}
    For the central $n_S=0.9649$ value from the Planck TT, TE, EE+lowE+lensing data~\cite{Planck:2018jri}, Eq.~(\ref{eq:Lambda}) gives $2.7331\times 10^{-3}\leq\Lambda/M_p\leq5.1006\times10^{-3}$ for the $15\leq\Delta\leq28$ range. The corresponding range of tensor-to-scalar ratio is $0.0035\leq r \leq 0.036$, which is compatible with the latest constraint on $r_{0.05}<0.036$ at $95\%$ confidence~\cite{BICEP:2021xfz}.  
    
Substituting Eq.~(\ref{eq:Lambda}) into Eq.~(\ref{eq:Delta}), we plot the parameter space of $\gamma$ and $M^2/(\beta M_p^2)$ (or effectively $\beta$) in the right panel of Fig.~\ref{fig:nsvsr}, where we adopt the $n_S=0.9649$ and $0.3<\log_{10}(f/M_p) <2.5$ from the Planck paper~\cite{Planck:2018jri} and vary $\Delta$ within the range of $15\leq\Delta\leq28$. The dark- and light-blue regions are plotted for $\Delta=15$ and $\Delta=28$, respectively. The lower (upper) limits of $f/M_p$ for different values of $\Delta$ are presented by solid (dashed) lines in the figure. When $\gamma=1$, the $\Delta$ becomes zero, which explains the divergence at $\gamma=1$. In this work, we consider $\gamma\neq1$ hence the divergence is irrelevant for our study. The left panel of Fig.~\ref{fig:nsvsr} shows that the CMB data prefer the positive $\Delta$ values; therefore, a combination $\beta(\gamma-1)$ in Eq.~(\ref{eq:Delta}) must also be positive. This implies $\beta>0$ for $\gamma>1$ and $\beta<0$ for $\gamma<1$, respectively. It is worth emphasizing here that the sufficient condition $H^2/M_p^2\ll1$ to avoid quantum gravity reads $M^2/[3\beta(\gamma-1)M_p^2]\ll1$ in our work and is satisfied for the parameter space presented in the figure. Thus, we conclude from Fig.~\ref{fig:nsvsr} that the natural inflation model is now consistent with the latest observational data for a broad range of model parameters. In the following section, to put further constraints to the inflationary predictions for natural inflation, we impose reheating considerations after inflation.

\section{Reheating constraints}\label{sec:reheating}
Inflation ends when equation-of-state parameter becomes $\omega_{\text{inf}}=-1/3$. The standard inflationary cosmology then assumes that, after inflation, the universe undergoes a phase of reheating, during which the inflaton field coherently oscillates at the minimum of its potential, decaying and transferring its energy to a relativistic plasma, and populates the universe with ordinary matter. As for the reheating phase, we follow the approaches proposed in Ref.~\cite{Martin:2010kz, Adshead:2010mc, Mielczarek:2010ag, Dodelson:2003vq, Easther:2011yq, Dai:2014jja, Munoz:2014eqa,Creminelli:2014fca, Cai:2015soa}, especially~\cite{Dai:2014jja, Munoz:2014eqa}, in which the present day observations are related to the evolution of the inflaton field during inflation. The key is the relation between the comoving scale $k$ today and that during inflation. In other words, the comoving Hubble radius, $a_k H_k=k$, exiting the horizon is associated with that of present time by
\begin{align}\label{eq:logk1}
    \ln\frac{k}{a_0 H_0} = \ln\frac{(a H)_{k}}{(aH)_{e}}+\ln\frac{(a H)_{e}}{(a H)_{re}}+\ln\frac{(a H)_{re}}{(aH)_{eq}} + \ln\frac{(a H)_{eq}}{a_0 H_0}\,.
\end{align}
The number of $e$-folds during each inflation, reheating, and radiation dominated epoch; $N_k=\ln(a_{e}/a_{k})$, $N_{re}=\ln(a_{re}/a_{e})$, and $N_{RD}=\ln(a_{eq}/a_{re})$, respectively, can be used to simplify Eq.~(\ref{eq:logk1}) as
\begin{equation}\label{eq:logk2}
\ln\frac{k}{a_0H_0}=-N_k-N_{re}-N_{RD}+\ln\frac{a_{eq}}{a_0}+\ln\frac{H_k}{H_{0}}\,,
\end{equation} 
where ``$a_\text{eq}$'' is the scale factor at the matter and radiation equality, and the subscript ``0'' denotes the present day value of each quantity. 
The Hubble parameter during inflation is obtained from Eq.~(\ref{eq:powerT}) with an assumption $\mathcal{P}_T(k_\ast)=r \mathcal{P}_S(k_\ast)$ to be $H_k=\sqrt{2}\pi M_p(r\mathcal{P}_S)^{1/2}c_T^{3/2}$, where $c_T\simeq1$ for our case. The ratio between the energy density $\rho_{e}$ at the end of the inflation and the energy density $\rho_{re}$ at the end of the reheating depends on the equation-of-state parameter and the duration of reheating and is 
\begin{equation}\label{eq:EoS}
\frac{\rho_{re}}{\rho_{e}}=e^{-3(1+\omega_{re})N_{re}}\,.
\end{equation} 
The energy density $\rho_{e}$ at the end of inflation is then computed from the modified Friedmann equations Eq.~(\ref{eq:MFEq}) to be $\rho_e\simeq V(\phi_e)$. 
Assuming the conservation of entropy and the current neutrino temperature, $T_{\nu,0}=(4/11)^{1/3}T_0$, the energy density at reheating is
\begin{equation}\label{eq:rre}
\rho_{re}=\frac{\pi^2g_{re}}{30}T_{re}^4\,,
\end{equation}
where
\begin{equation}\label{eq:Tre}
T_{re}=\left(\frac{43}{11g_{s,re}}\right)^{\frac{1}{3}}\frac{a_0}{a_{re}} T_0\,.
\end{equation} 
The number of $e$-folds during reheating can be obtained from Eq.~(\ref{eq:EoS}) with Eq.~(\ref{eq:rre}) and Eq.~(\ref{eq:logk2}) as
\begin{align}\label{eq:Nre}
N_{re}=\frac{4}{1-3\omega_{re}}\left[-N_k-\ln\frac{k}{a_0T_0}-\frac{1}{4}\ln\frac{30}{\pi^2g_{re}} 
-\frac{1}{3}\ln\frac{11g_{s,re}}{43}
-\frac{1}{4}\ln V(\phi_e)+\frac{1}{2}\ln\left(2\pi^2 M_p^2r\mathcal{P}_S\right)\right]\,.
\end{align}
The corresponding reheating temperature, $T_{re}$, is
\begin{align}\label{eq:Tre01}
T_{re}^4=\left(\frac{30}{\pi^2g_{re}}\right) 
V(\phi_e)e^{-3(1+\omega_{re})N_{re}}\,.
\end{align}

\begin{figure}
    \centering
    \includegraphics[width=0.45\textwidth]{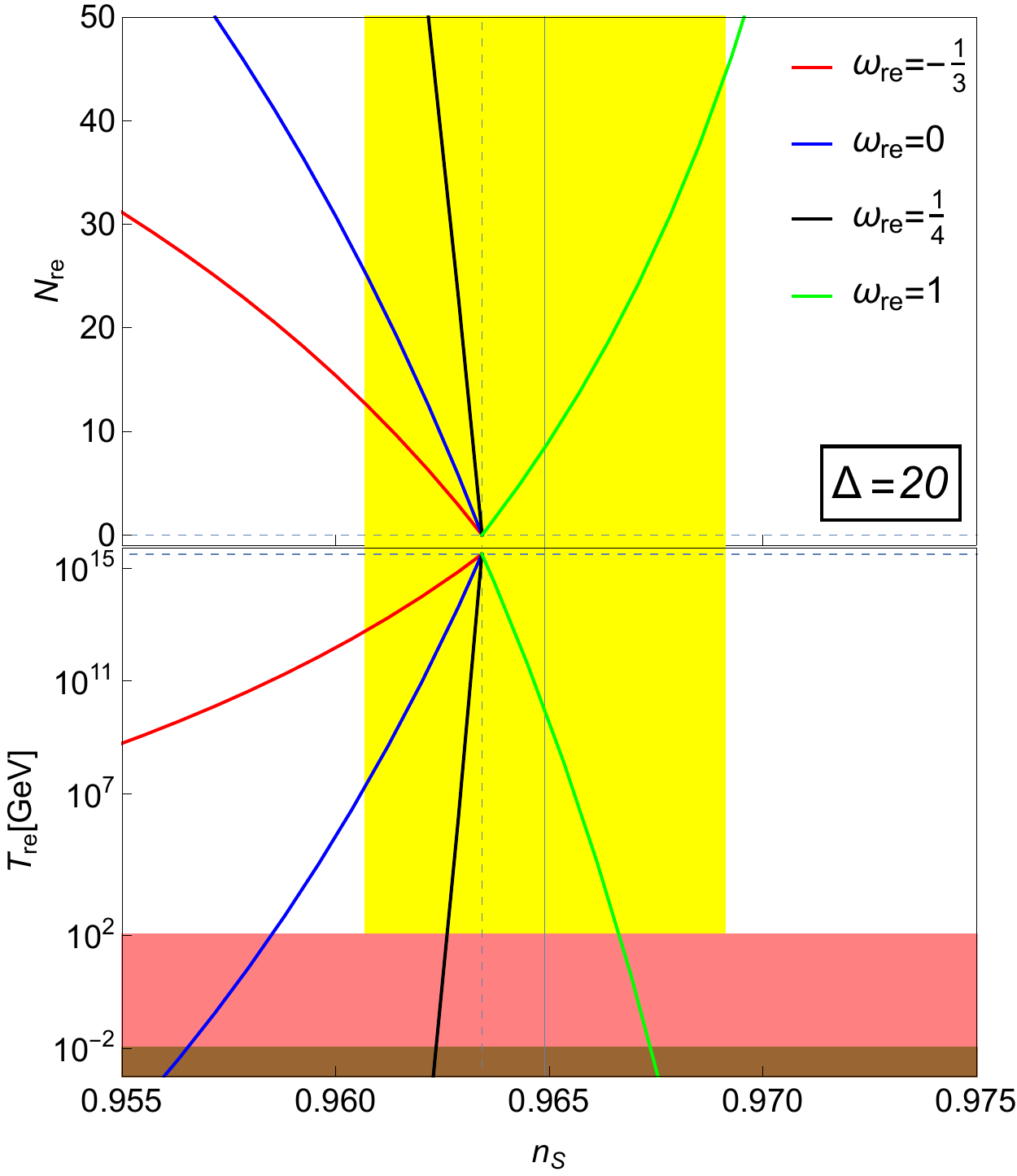}\hspace{0.5 cm}
    \includegraphics[width=0.45\textwidth]{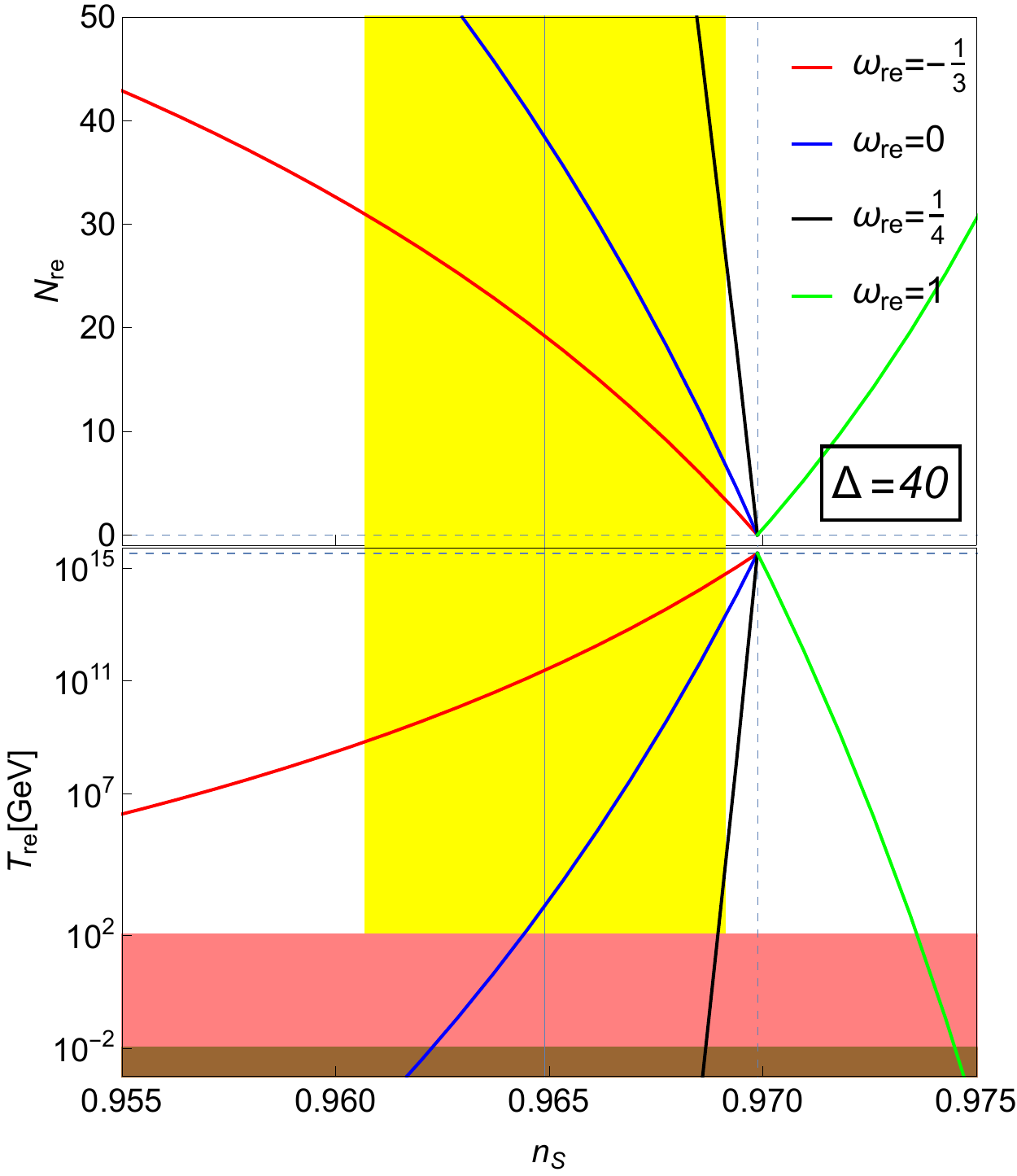}
    \caption{The $n_S-T_{re}$ (bottom) and $n_S-N_{re}$ (top) plots for natural inflation with potential given in Eq.~(\ref{eq:potential}). The yellow shaded region indicates the current $1\sigma$ range of $n_S=0.9649\pm0.0042$ ($68\%, TT, TE, EE+lowE+lensing$) from Planck data~\cite{Planck:2018jri}. The pink and brown shaded regions indicate the $T_{re}\leq100$GeV and $T_{re}\leq10$MeV energy scales, respectively. The $\omega_{re}$ values residing inside the shaded regions are favored. }
    \label{fig:Trevsns}
\end{figure}

Assuming the effective equation-of-state parameter $\omega_{re}$ is functionally constant, we plot the temperature $T_{re}$ and the duration of reheating as functions of the spectral index $n_S$ in Fig.~\ref{fig:Trevsns}. The left and right panels correspond to two different values of $\Delta$. In each figure, we choose four different values of $\omega_{re}$; namely, $\omega_{re}=-1/3$ (red), $\omega_{re}=0$ (blue), $\omega_{re}=0.25$ (black), and $\omega_{re}=1$ (green). The smallest $\omega_{re}=-1/3$ value is required for inflation end, while the largest $\omega_{re}=1$ one, the most conservative upper limit, comes from the causality. Thus, the $\omega_{re}$ value varies within the interval of $-1/3\leq\omega_{re}\leq1$. The background yellow shaded region presents the current $1\sigma$ range of $n_S=0.9649\pm0.0042$ ($68\%, TT, TE, EE+lowE+lensing$) from Planck data~\cite{Planck:2018jri}. The pink and brown shaded regions with $T_{re}\leq100$GeV and $T_{re}\leq10$MeV indicate the constraints on energy scales the electroweak and the big-bang nucleosynthesis, respectively. The $\omega_{re}$ values residing inside the shaded regions are favored. The curves of all $\omega_{re}$ intersect at $(n_S, T_{re}^{max}) = (0.9634, 3.04\times10^{15} \text{GeV})$ for $\Delta=20$ and $(n_S, T_{re}^{max}) = (0.9698, 3.26\times10^{15} \text{GeV})$ for $\Delta=40$ and indicate the instantaneous reheating ($N_{re}=0$) for our model. The instantaneous reheating temperature $T^{max}_{re}$ and corresponding spectral index $n_S$ values slightly increase as the $\Delta$ value increases, which is due to the potential energy value at the end of inflation depends on the value of $\Delta$, see Eqs.~(\ref{eq:potphie}) and (\ref{eq:Tre01}).  By matching the lines of the same colors in both upper and lower panels, one can estimate the duration and corresponding temperature of reheating. Thus, for given $\omega_{re}$, the figure shows that the longer the reheating lasts, the lower the temperature gets at the end of reheating. For the increasing direction of $n_S$, the $T_{re}$ increases, while the $N_{re}$ decreases, toward the intersecting point for $\omega_{re}<1/3$. As for the case of $\omega_{re}>1/3$, the $T_{re}$ decreases, while $N_{re}$ increases, away from the intersecting point of all lines.
\begin{figure}
    \centering
    \includegraphics[width=0.45\textwidth]{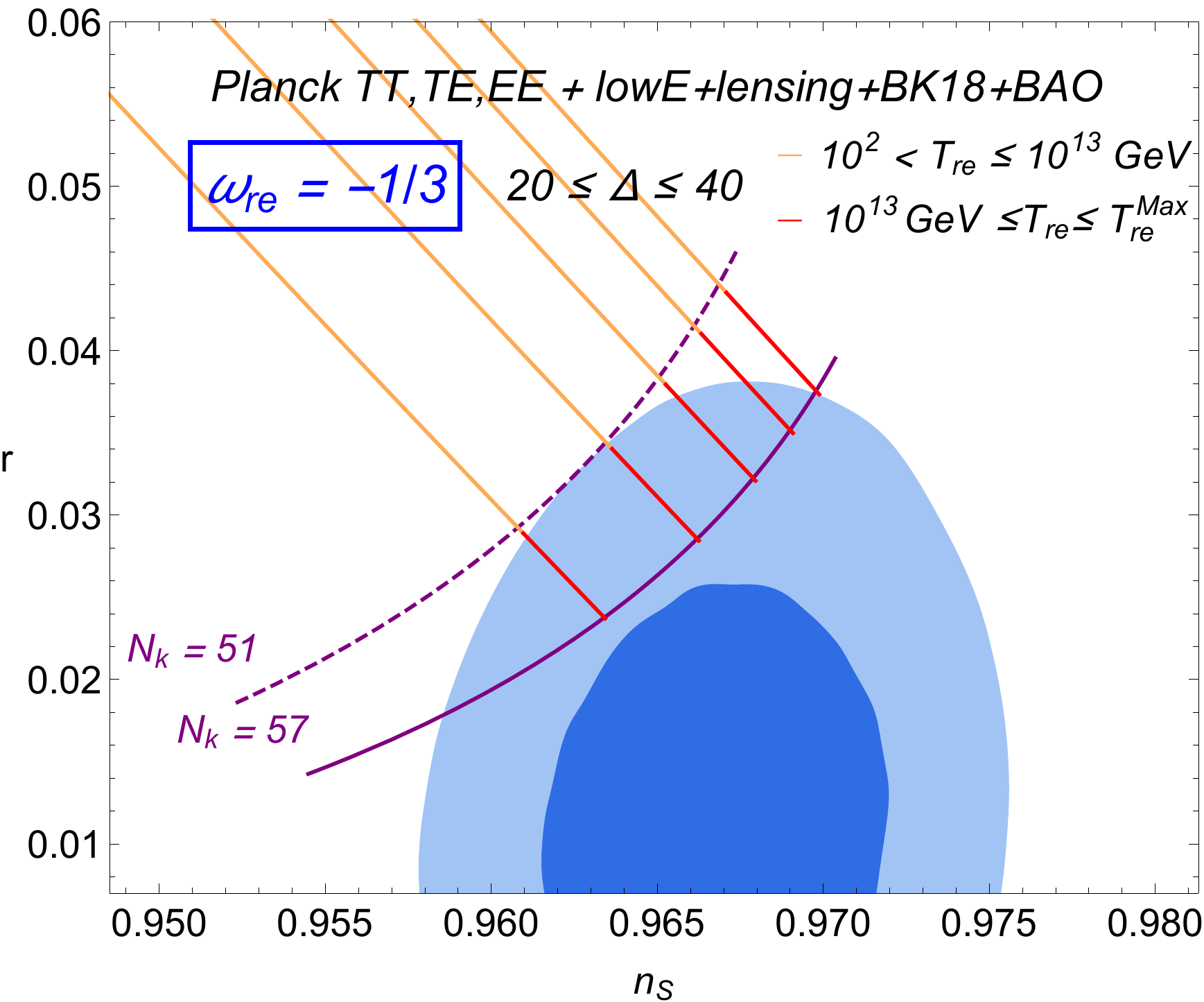} \hspace{0.5 cm}
    \includegraphics[width=0.45\textwidth]{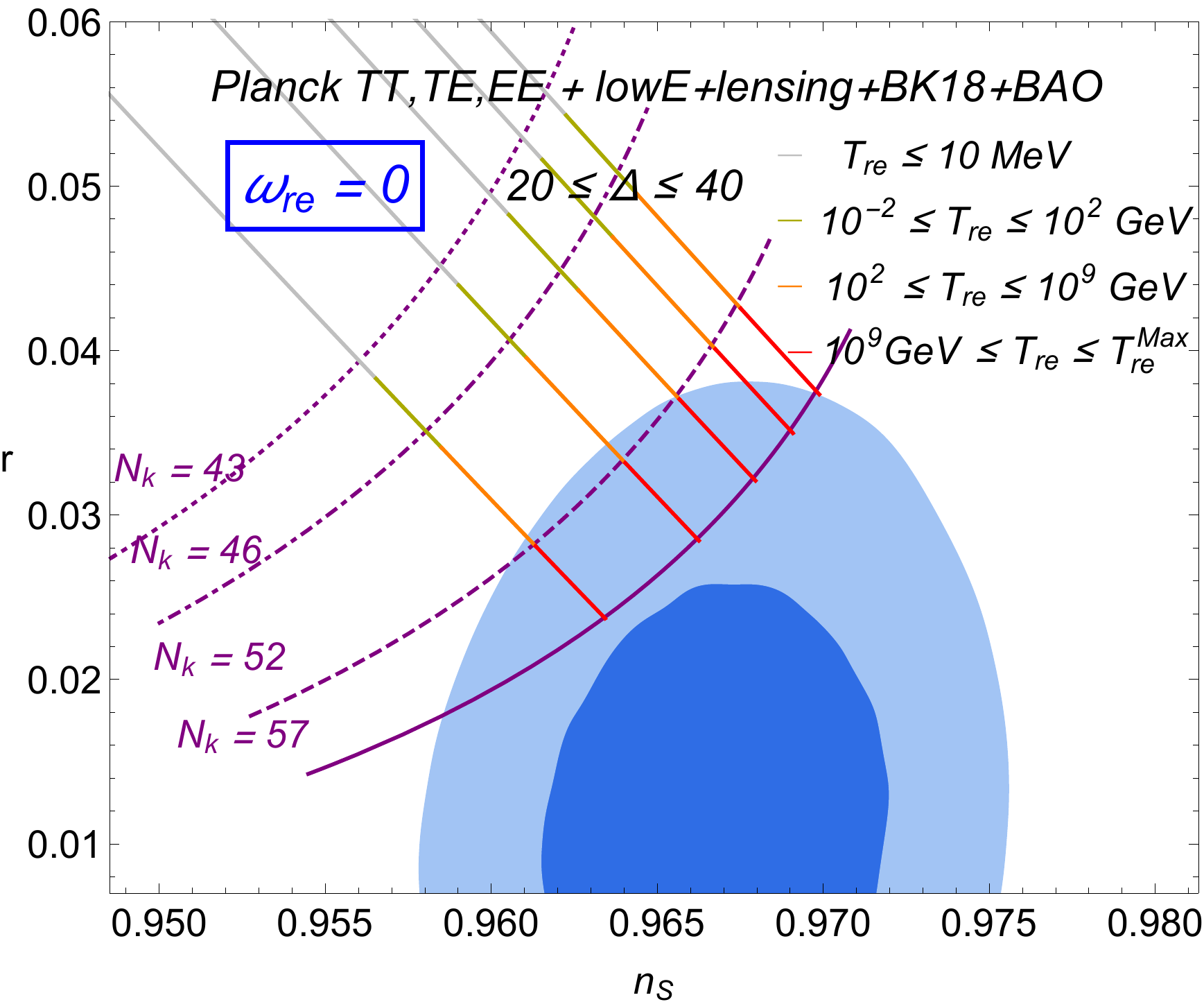}\\
    \includegraphics[width=0.45\textwidth]{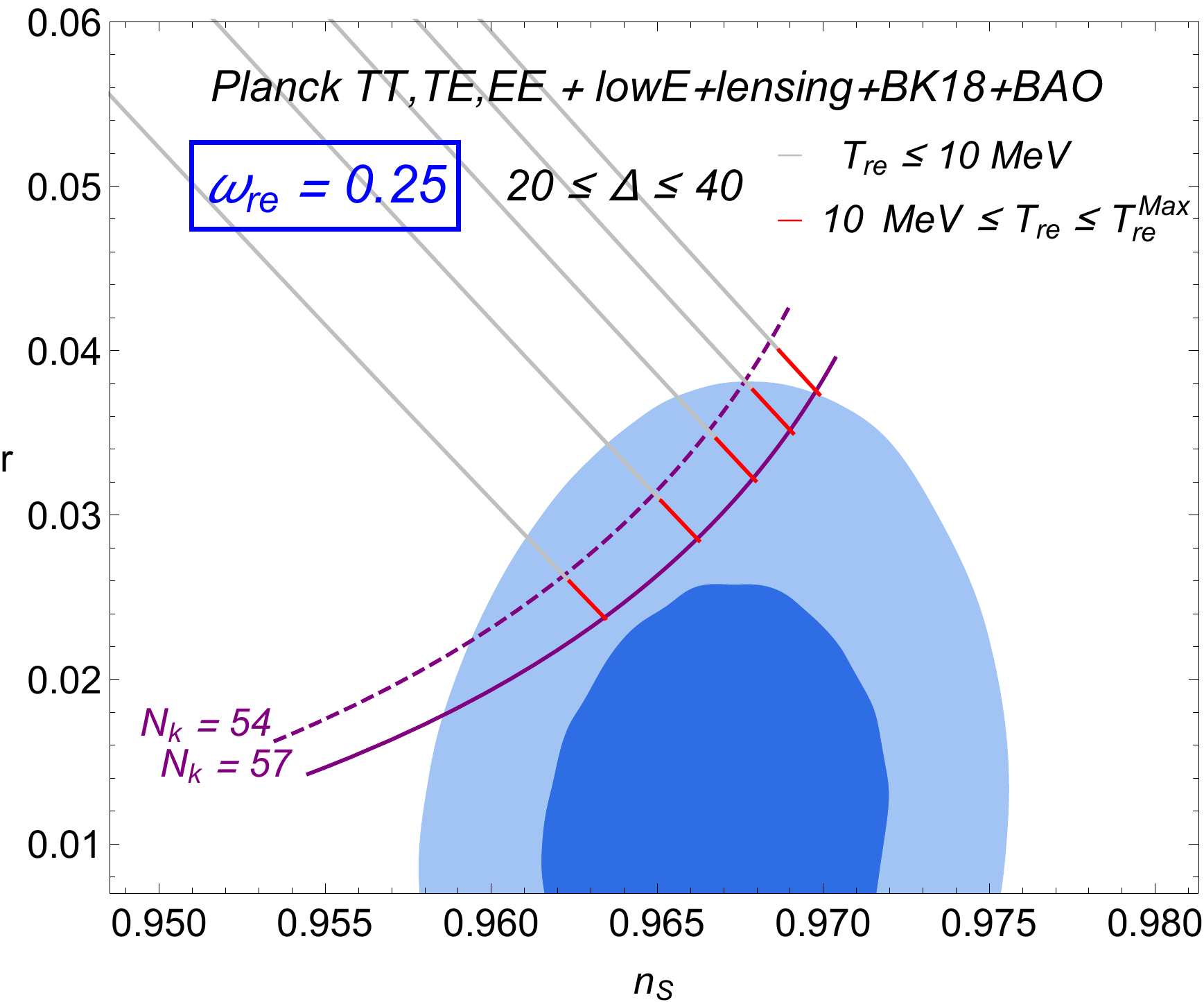} \hspace{0.5 cm}
    \includegraphics[width=0.45\textwidth]{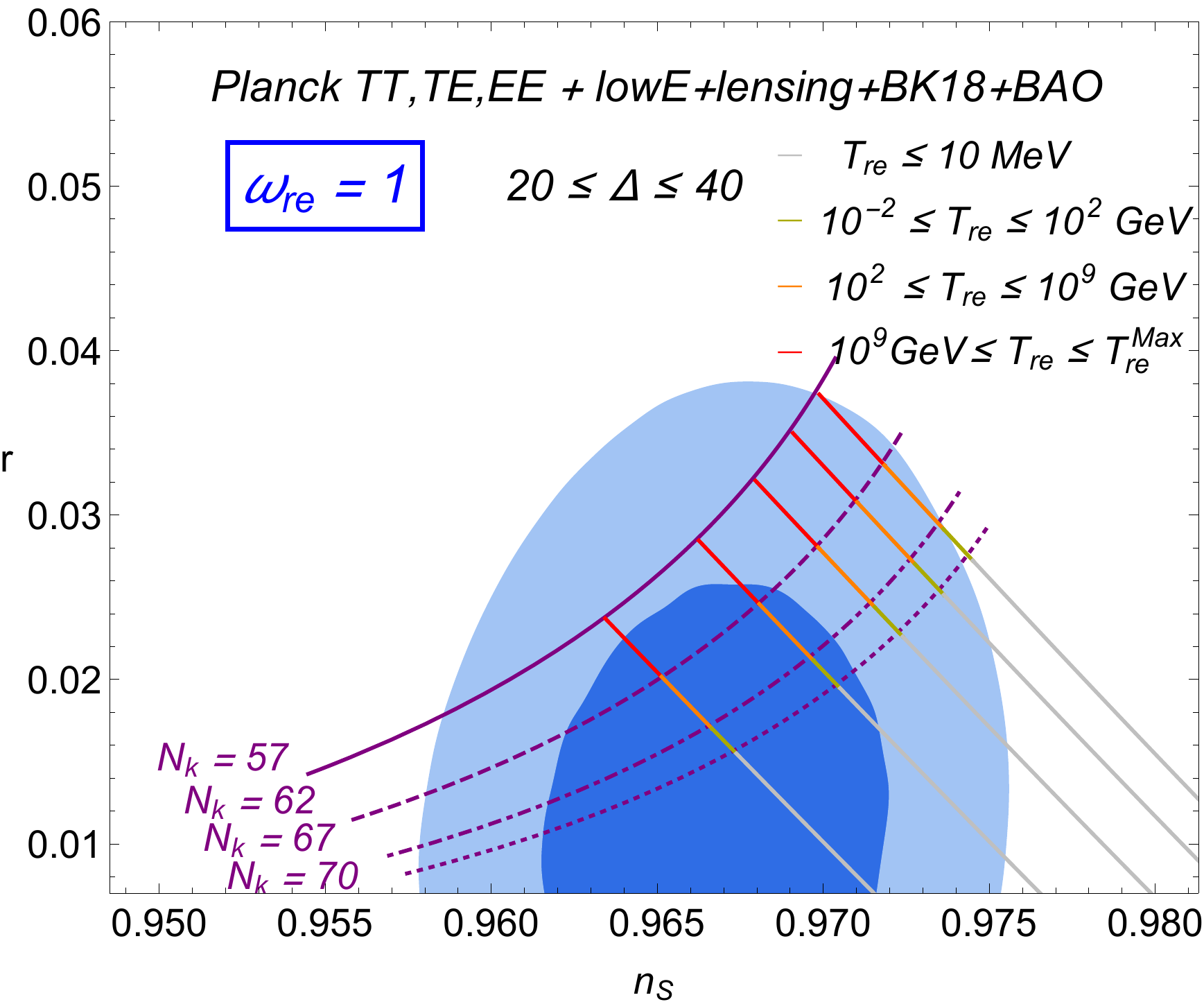} 
    \caption{The $n_S-r$ plot for natural inflation with potential given in Eq.~(\ref{eq:potential}). The shaded regions are the same as Fig.~\ref{fig:nsvsr}.}
    \label{fig:fignsr}
\end{figure}

In Eq.~(\ref{eq:Nkns}), the duration $N_k$ of inflation is expressed in terms of $n_S$ and $\Delta$. Also, the $n_S$ can be expressed in terms of reheating parameters, including $\omega_{re}, N_{re}$, and $T_{re}$. Thus, by taking the reheating considerations into account, we plot $n_S-r$ predictions once again in Fig.~\ref{fig:fignsr} for the same values of $\omega_{re}$ as Fig.~\ref{fig:Trevsns}. Since $N_{re}$ and $T_{re}$ are related to each other, we choose $T_{re}$ as another reheating parameter in Fig.~\ref{fig:Trevsns}. As it is in Fig.~\ref{fig:nsvsr}, the underlying diagonal gray lines correspond to the different values of $\Delta$, and it increases from bottom left to top right within the interval of $20\leq\Delta\leq 40$. The purple lines are some reference values of $N_k$. As the labels indicate in each panel, the different colors on the diagonal lines correspond to the different values of $T_{re}$. The main implication of this figure is that both the $\omega_{re}$ and the $T_{re}$ at the end of reheating provide us with information about how long inflation had to be lasted to be compatible with the CMB observations. For example, if $\omega_{re}<1/3$, inflation cannot last longer than $N_k\simeq57$ $e$-folds due to our reheating considerations. Thus, the prediction of $n_S$ and $r$ is only $2\sigma$ consistent with the observations. Conversely, the duration of inflation must be longer than $N_k\gtrsim57$ for $\omega_{re}>1/3$. As a result, the theoretical prediction of our model is still consistent with the CMB data, the $n_S$ and $r$ values residing inside $1\sigma$ contour. The turning point is the instantaneous reheating with $N_{re}=0$, which also indicates $\omega_{re}=1/3$. Since reheating must have occurred instantaneously after inflation, the estimated duration of inflation is $N_k\simeq57$ $e$-folds, the exact value is slightly different for different $\Delta$ values as the figure shows. Fig.~\ref{fig:fignsr} also shows that the CMB data prefer different ranges of reheating temperature depending on the values of $\omega_{re}$. For equation of state approaching $\omega_{re}\simeq 1/3$, the reheating temperature from as low as a few MeV to as large as $10^{15}$GeV can be achieved.

\section{Conclusions} \label{sec:conclusions}

For a cosmological model described by Eq.~(\ref{eq:action}), where the derivative self-interaction of a scalar field and its kinetic coupling with gravity is presented, we placed observational constraints on the natural inflation model. Then, we investigated the effects of reheating on the inflationary predictions. The interactions we introduced in this work are assumed to give equally important contributions during inflation such that $|\gamma|\sim\mathcal{O}(1)$ in Eq.~(\ref{eq:gamma}). The main analytic results of inflationary predictions for our model are obtained in Eqs.~(\ref{eq:powerS})--(\ref{eq:rofphi}), where the presence of $\mathcal{A}$ indicates the non-zero effects of the interactions mentioned above. These equations do not depend on details of inflaton potential. To give a sizable effect during inflation, deviating from the standard single-field inflation model in Einstein gravity, the $\mathcal{A}$ should be $\mathcal{A}\gtrsim1$. Eq.~(\ref{eq:MFEq}) shows that the additional interactions in our model significantly enhance the gravitational friction. Consequently, the tensor-to-scalar ratio is suppressed by a factor of $1+\mathcal{A}$ in Eq.~(\ref{eq:rofphi}). The suppression is more significant if $\mathcal{A}\gg1$; hence, the inflationary predictions can be compatible with the observations.    

In Sec.~\ref{sec:obsNI}, in light of the latest observational data~\cite{BICEP:2021xfz, Planck:2018jri}, we discussed the theoretical predictions of the natural inflation model with potential given in Eq.~(\ref{eq:potential}). Although we have several free parameters, the observable quantities of natural inflation depend only on a single parameter $\Delta$ as defined in Eq.~(\ref{eq:Delta}). In the $n_S-r$ plane in Fig.~\ref{fig:nsvsr}, we showed that the natural inflation model is now compatible with the observational data for a certain parameter range, mainly due to the suppressed tensor-to-scalar ratio $r$. For the predictions to be consistent with data at $1\sigma$ ($68\%$ confidence) level, natural inflation is supposed to last slightly longer ($N_k\gtrsim60$) than usually assumed. The right panel of Fig.~\ref{fig:nsvsr} shows the parameter space of $M$ and $\gamma$ parameters that provides the right value for the amplitude of the CMB spectrum. 

Imposing the effective equation-of-state $\omega_{re}$ parameter to be constant after inflation, we estimated the duration and the temperature of reheating in terms of inflationary parameters in Eqs.~(\ref{eq:Nre}) and~(\ref{eq:Tre01}), respectively. In Fig.~\ref{fig:Trevsns}, we plotted Eqs.~(\ref{eq:Nre}) and~(\ref{eq:Tre01}) as a function of $n_S$ for given $\Delta$ and $\omega_{re}$ values. The figure shows that the broad ranges of $\omega_{re}$ and $\Delta$ give compatible results with the CMB predictions on $n_S$. The combined result with Fig.~\ref{fig:fignsr} implies that, if $\omega_{re}>1/3$, inflation can last long enough$-$\emph{i.e.}, having $N_\ast\gtrsim60$. Thus, it is more compatible with the CMB data, the predictions residing inside the $1\sigma$ contour, see Fig.~\ref{fig:fignsr} for $\omega_{re}=1$. For the $\omega_{re}<1/3$ cases, the maximum number of $e$-folds in natural inflation is $N_k\simeq57$; hence the predictions reside inside the $2 \sigma$ contour only, see Fig.~\ref{fig:fignsr} for $\omega_{re}=-1/3,\,0,$ and $0.25$ cases. Our result also shows that the broad range of reheating temperatures, from as low as $\sim\mathcal{O}(10)$MeV to as large as $\sim\mathcal{O}(10^{15})$GeV, can be achieved at the end of reheating if the equation of state is closer to $1/3$. The maximum reheating temperature is $T_{re}^{max}\simeq 3\times 10^{15}$GeV, corresponding to an instantaneous reheating scenario for which $\omega_{re}=1/3$ and $N_{re}=0$.  

We conclude that the effects of reheating to inflationary predictions are important to tighten the parameter space of observable quantities, breaking degeneracy between inflationary predictions that otherwise overlap in the $n_S-r$ plane. We also found it is interesting to investigate the phenomenological details of reheating for the $\omega_{re}\gtrsim1/3$. According to our result, in that case, inflation can last longer than usually assumed; hence more consistent with the CMB data, and the broad range of reheating temperature is also accessible. 
\section*{Acknowledgements} \label{sec:acknowledgements}
    CC is supported by the Leung Center for Cosmology and Particle Astrophysics (LeCosPA), National Taiwan University. SK and GT are supported by the National Research Foundation of Korea (NRF-2016R1D1A1B04932574, NRF-2021R1A2C1005748). We are pleased to appreciate Asian Pacific Center for Theoretical Physics (APCTP) for its hospitality during completion of this work.


\begin{thebibliography}{99}
\bibitem{Guth:1980zm}
A.~H.~Guth,
``The Inflationary Universe: A Possible Solution to the Horizon and Flatness Problems,''
Phys. Rev. D \textbf{23}, 347-356 (1981);

\bibitem{Linde:1981mu}
A.~D.~Linde,
``A New Inflationary Universe Scenario: A Possible Solution of the Horizon, Flatness, Homogeneity, Isotropy and Primordial Monopole Problems,''
Phys. Lett. B \textbf{108}, 389-393 (1982);

\bibitem{Linde:1982zj}
A.~D.~Linde,
``Coleman-Weinberg Theory and a New Inflationary Universe Scenario,''
Phys. Lett. B \textbf{114}, 431-435 (1982);

\bibitem{Albrecht:1982wi}
A.~Albrecht and P.~J.~Steinhardt,
``Cosmology for Grand Unified Theories with Radiatively Induced Symmetry Breaking,''
Phys. Rev. Lett. \textbf{48}, 1220-1223 (1982);

\bibitem{Linde:1983gd}
A.~D.~Linde,
``Chaotic Inflation,''
Phys. Lett. B \textbf{129}, 177-181 (1983);

\bibitem{Boomerang:2000efg}
P.~de Bernardis \textit{et al.} [Boomerang],
``A Flat universe from high resolution maps of the cosmic microwave background radiation,''
Nature \textbf{404}, 955-959 (2000);

\bibitem{SDSS:2003eyi}
M.~Tegmark \textit{et al.} [SDSS],
``Cosmological parameters from SDSS and WMAP,''
Phys. Rev. D \textbf{69}, 103501 (2004);

\bibitem{SDSS:2004kqt}
U.~Seljak \textit{et al.} [SDSS],
``Cosmological parameter analysis including SDSS Ly-alpha forest and galaxy bias: Constraints on the primordial spectrum of fluctuations, neutrino mass, and dark energy,''
Phys. Rev. D \textbf{71}, 103515 (2005);

\bibitem{Blake:2011en}
C.~Blake, E.~Kazin, F.~Beutler, T.~Davis, D.~Parkinson, S.~Brough, M.~Colless, C.~Contreras, W.~Couch and S.~Croom, \textit{et al.}
``The WiggleZ Dark Energy Survey: mapping the distance-redshift relation with baryon acoustic oscillations,''
Mon. Not. Roy. Astron. Soc. \textbf{418}, 1707-1724 (2011);

\bibitem{Planck:2018jri}
Y.~Akrami \textit{et al.} [Planck],
``Planck 2018 results. X. Constraints on inflation,''
Astron. Astrophys. \textbf{641}, A10 (2020);

P.~A.~R.~Ade \textit{et al.} [Planck],
``Planck 2015 results. XX. Constraints on inflation,''
Astron. Astrophys. \textbf{594}, A20 (2016);

\bibitem{Freese:1990rb}
K.~Freese, J.~A.~Frieman and A.~V.~Olinto,
``Natural inflation with pseudo - Nambu-Goldstone bosons,''
Phys. Rev. Lett. \textbf{65}, 3233-3236 (1990);

\bibitem{Adams:1992bn}
F.~C.~Adams, J.~R.~Bond, K.~Freese, J.~A.~Frieman and A.~V.~Olinto,
``Natural inflation: Particle physics models, power law spectra for large scale structure, and constraints from COBE,''
Phys. Rev. D \textbf{47}, 426-455 (1993);

\bibitem{Freese:2004un}
K.~Freese and W.~H.~Kinney,
``On: Natural inflation,''
Phys. Rev. D \textbf{70}, 083512 (2004);

\bibitem{Savage:2006tr}
C.~Savage, K.~Freese and W.~H.~Kinney,
``Natural Inflation: Status after WMAP 3-year data,''
Phys. Rev. D \textbf{74}, 123511 (2006);

\bibitem{Freese:2014nla}
K.~Freese and W.~H.~Kinney,
``Natural Inflation: Consistency with Cosmic Microwave Background Observations of Planck and BICEP2,''
JCAP \textbf{03}, 044 (2015);

\bibitem{Freese:1994fp}
K.~Freese,
``A Coupling of pseudoNambu-Goldstone bosons to other scalars and role in double field inflation,''
Phys. Rev. D \textbf{50}, 7731-7734 (1994);

\bibitem{Kawasaki:2000yn}
M.~Kawasaki, M.~Yamaguchi and T.~Yanagida,
``Natural chaotic inflation in supergravity,''
Phys. Rev. Lett. \textbf{85}, 3572-3575 (2000);

\bibitem{Arkani-Hamed:2003xts}
N.~Arkani-Hamed, H.~C.~Cheng, P.~Creminelli and L.~Randall,
``Extra natural inflation,''
Phys. Rev. Lett. \textbf{90}, 221302 (2003);
N.~Arkani-Hamed, H.~C.~Cheng, P.~Creminelli and L.~Randall,
``Pseudonatural inflation,''
JCAP \textbf{07}, 003 (2003);

\bibitem{Kaplan:2003aj}
D.~E.~Kaplan and N.~J.~Weiner,
``Little inflatons and gauge inflation,''
JCAP \textbf{02}, 005 (2004);

\bibitem{Firouzjahi:2003zy}
H.~Firouzjahi and S.~H.~H.~Tye,
``Closer towards inflation in string theory,''
Phys. Lett. B \textbf{584}, 147-154 (2004);

\bibitem{Hsu:2004hi}
J.~P.~Hsu and R.~Kallosh,
``Volume stabilization and the origin of the inflaton shift symmetry in string theory,''
JHEP \textbf{04}, 042 (2004);

\bibitem{Kim:2004rp}
J.~E.~Kim, H.~P.~Nilles and M.~Peloso,
``Completing natural inflation,''
JCAP \textbf{01}, 005 (2005);

\bibitem{Huang:2015cke}
Q.~G.~Huang, K.~Wang and S.~Wang,
``Inflation model constraints from data released in 2015,''
Phys. Rev. D \textbf{93}, no.10, 103516 (2016);

\bibitem{Stein:2021uge}
N.~K.~Stein and W.~H.~Kinney,
``Natural Inflation After Planck 2018,''
[arXiv:2106.02089 [astro-ph.CO]];

\bibitem{BICEP:2021xfz}
P.~A.~R.~Ade \textit{et al.} [BICEP and Keck],
``Improved Constraints on Primordial Gravitational Waves using Planck, WMAP, and BICEP/Keck Observations through the 2018 Observing Season,''
Phys. Rev. Lett. \textbf{127} (2021) no.15, 151301;

\bibitem{Munoz:2014eqa}
J.~B.~Munoz and M.~Kamionkowski,
``Equation-of-State Parameter for Reheating,''
Phys. Rev. D \textbf{91}, no.4, 043521 (2015);

\bibitem{Zhang:2018wbn}
N.~Zhang, Y.~B.~Wu, J.~W.~Lu, C.~W.~Sun, L.~J.~Shou and H.~Z.~Xu,
``Constraints on the generalized natural inflation after Planck 2018,''
Chin. Phys. C \textbf{44}, no.9, 095107 (2020);

\bibitem{Civiletti:2020fkm}
M.~Civiletti and B.~Delacruz,
``Natural inflation with natural number of $e$-foldings,''
Phys. Rev. D \textbf{101}, no.4, 043534 (2020);

\bibitem{Forconi:2021que}
M.~Forconi, W.~Giar\`e, E.~Di Valentino and A.~Melchiorri,
``Cosmological constraints on slow roll inflation: An update,''
Phys. Rev. D \textbf{104}, no.10, 103528 (2021);


\bibitem{Tumurtushaa:2019bmc}
G.~Tumurtushaa,
``Inflation with Derivative Self-interaction and Coupling to Gravity,''
Eur. Phys. J. C \textbf{79}, no.11, 920 (2019);

\bibitem{Bayarsaikhan:2020jww}
B.~Bayarsaikhan, S.~Koh, E.~Tsedenbaljir and G.~Tumurtushaa,
``Constraints on dark energy models from the Horndeski theory,''
JCAP \textbf{11}, 057 (2020);

\bibitem{Chen:2021nio}
P.~Chen, S.~Koh and G.~Tumurtushaa,
``Primordial black holes and induced gravitational waves from inflation in the Horndeski theory of gravity,''
[arXiv:2107.08638 [gr-qc]];

\bibitem{Kobayashi:2011nu}
T.~Kobayashi, M.~Yamaguchi and J.~Yokoyama,
``Generalized G-inflation: Inflation with the most general second-order field equations,''
Prog. Theor. Phys. \textbf{126}, 511-529 (2011);

\bibitem{Horndeski:1974wa}
G.~W.~Horndeski,
``Second-order scalar-tensor field equations in a four-dimensional space,''
Int. J. Theor. Phys. \textbf{10}, 363-384 (1974);

\bibitem{Deffayet:2011gz}
C.~Deffayet, X.~Gao, D.~A.~Steer and G.~Zahariade,
``From k-essence to generalised Galileons,''
Phys. Rev. D \textbf{84}, 064039 (2011);


\bibitem{Ostrogradski}
M. Ostrogradski, Mem. Ac. St. Petersbourg VI 4, 385 (1850);

\bibitem{Kobayashi:2019hrl}
T.~Kobayashi,
``Horndeski theory and beyond: a review,''
Rept. Prog. Phys. \textbf{82}, no.8, 086901 (2019);

\bibitem{Germani:2010gm}
C.~Germani and A.~Kehagias,
``New Model of Inflation with Non-minimal Derivative Coupling of Standard Model Higgs Boson to Gravity,''
Phys. Rev. Lett. \textbf{105}, 011302 (2010);

C.~Germani and A.~Kehagias,
``UV-Protected Inflation,''
Phys. Rev. Lett. \textbf{106}, 161302 (2011);

\bibitem{Tsujikawa:2012mk}
S.~Tsujikawa,
``Observational tests of inflation with a field derivative coupling to gravity,''
Phys. Rev. D \textbf{85}, 083518 (2012);

\bibitem{Tsujikawa:2013ila}
S.~Tsujikawa, J.~Ohashi, S.~Kuroyanagi and A.~De Felice,
``Planck constraints on single-field inflation,''
Phys. Rev. D \textbf{88}, no.2, 023529 (2013);

\bibitem{Germani:2011ua}
C.~Germani and Y.~Watanabe,
``UV-protected (Natural) Inflation: Primordial Fluctuations and non-Gaussian Features,''
JCAP \textbf{07}, 031 (2011);

\bibitem{Kamada:2010qe}
K.~Kamada, T.~Kobayashi, M.~Yamaguchi and J.~Yokoyama,
``Higgs G-inflation,''
Phys. Rev. D \textbf{83}, 083515 (2011);

\bibitem{Abbott:1982hn}
L.~F.~Abbott, E.~Farhi and M.~B.~Wise,
``Particle Production in the New Inflationary Cosmology,''
Phys. Lett. B \textbf{117}, 29 (1982);

\bibitem{Dolgov:1982th}
A.~D.~Dolgov and A.~D.~Linde,
``Baryon Asymmetry in Inflationary Universe,''
Phys. Lett. B \textbf{116}, 329 (1982);

\bibitem{Albrecht:1982mp}
A.~Albrecht, P.~J.~Steinhardt, M.~S.~Turner and F.~Wilczek,
``Reheating an Inflationary Universe,''
Phys. Rev. Lett. \textbf{48}, 1437 (1982);

\bibitem{Allahverdi:2010xz}
R.~Allahverdi, R.~Brandenberger, F.~Y.~Cyr-Racine and A.~Mazumdar,
``Reheating in Inflationary Cosmology: Theory and Applications,''
Ann. Rev. Nucl. Part. Sci. \textbf{60}, 27-51 (2010);

\bibitem{Dodelson:2003vq}
S.~Dodelson and L.~Hui,
``A Horizon ratio bound for inflationary fluctuations,''
Phys. Rev. Lett. \textbf{91}, 131301 (2003);

\bibitem{Martin:2010kz}
J.~Martin and C.~Ringeval,
``First CMB Constraints on the Inflationary Reheating Temperature,''
Phys. Rev. D \textbf{82}, 023511 (2010);

\bibitem{Adshead:2010mc}
P.~Adshead, R.~Easther, J.~Pritchard and A.~Loeb,
``Inflation and the Scale Dependent Spectral Index: Prospects and Strategies,''
JCAP \textbf{02}, 021 (2011);

\bibitem{Mielczarek:2010ag}
J.~Mielczarek,
``Reheating temperature from the CMB,''
Phys. Rev. D \textbf{83}, 023502 (2011);

\bibitem{Easther:2011yq}
R.~Easther and H.~V.~Peiris,
``Bayesian Analysis of Inflation II: Model Selection and Constraints on Reheating,''
Phys. Rev. D \textbf{85}, 103533 (2012);

\bibitem{Dai:2014jja}
L.~Dai, M.~Kamionkowski and J.~Wang,
``Reheating constraints to inflationary models,''
Phys. Rev. Lett. \textbf{113}, 041302 (2014);

\bibitem{Creminelli:2014fca}
P.~Creminelli, D.~L\'opez Nacir, M.~Simonovi\'c, G.~Trevisan and M.~Zaldarriaga,
``$\phi^2$ Inflation at its Endpoint,''
Phys. Rev. D \textbf{90}, no.8, 083513 (2014);

\bibitem{Cai:2015soa}
R.~G.~Cai, Z.~K.~Guo and S.~J.~Wang,
``Reheating phase diagram for single-field slow-roll inflationary models,''
Phys. Rev. D \textbf{92}, 063506 (2015).

















\end{thebibliography}
\end{document}